\documentclass[12pt]{article}
\usepackage{a4,amsmath}
\begin{document}

\begin{Large}
\noindent
{\bf Some physical displays of the space anisotropy relevant to the feasibility of its being detected at a laboratory}
\end{Large}

\vspace*{10mm}
\begin{large}
\noindent
{\bf George Yu. Bogoslovsky}
\end{large}

\vspace*{2mm}
\noindent
Skobeltsyn Institute of Nuclear Physics\\
Lomonosov Moscow State University\\
119991 Moscow, Russia.\\
E-mail: bogoslov@theory.sinp.msu.ru

\vspace*{15mm}
\begin{small}
\noindent
The impact of local space anisotropy on the transverse Doppler effect is examined. Two types of laboratory experiments aimed at seeking and measuring the local space anisotropy are proposed. In terms of the conventional special relativity theory, which treats 3D space to be locally isotropic, the experiments are of the type of ``null-experiments''. In the first-type experiments, a feasible Doppler shift of frequency is measured by the M\"ossbauer effect, with the M\"ossbauer source and absorber being located at two identical and diametrically opposed distances from the center of a rapidly rotating rotor, while the $\gamma$-quanta are recorded by two stationary and oppositely positioned proportional counters. Either of the counters records only those $\gamma$-quanta that passed through the absorber at the moment of the passage of the latter near a counter. The second-type experiments are made using the latest radio physics techniques for generating monochromatic oscillations and for recording weak signals. The effect expected due to space anisotropy consists in frequency modulation of the harmonic oscillations coming to a receiver that rotates at a constant velocity around the monochromatic wave emitter. In this case the modulation depth proves to be proportional to the space anisotropy magnitude.
\end{small}

\vspace*{15mm}
\noindent
{\large{\bf 1\,. Introduction}}

\vspace*{3mm}
\noindent
The recent rapid development of theoretical physics, astrophysics, and elementary particle physics has brought a much deeper insight into space physics. It has become clear, in particular, that the Riemannian model of locally isotropic space time can but approximately describe the properties of the real space.

As long ago as the early seventies of the former century, the researches into the extensive air showers of elementary particles induced by super high-energy cosmic ray protons yielded the pioneer experimental data indicating that the true behavior of the energy spectrum of primary cosmic protons contradicted the respective theoretical predictions [1],\,[2] made in terms of the locally isotropic space model. The recent studies have finally confirmed the discrepancy between the theoretical predictions  and the experimental data on the primary cosmic proton spectrum ( the absence of so-called GZK effect ). Besides, such phenomena were discovered as, for instance, the violation of discrete space-time symmetries in weak interactions of elementary particles, the temperature anisotropy of the microwave background radiation, and the offbeat behavior of the Hubble parameter that characterizes the Universe expansion. All the phenomena cannot be described in any way consistently in terms of the theory based on the model for locally isotropic space-time.

Generally speaking, the void space, or the space filled with locally isotropic matter, can only exhibit local isotropy. As to the void space, it is merely a mathematical idealization of physical vacuum that consists of locally isotropic fluctuations of quantized fields.
As to the locally isotropic matter, this can be exemplified by the Higgs condensate, i.e. the classical constant scalar field induced by spontaneous violation of gauge symmetry in the unified theory for electroweak interactions. The standard version of the theory describes quite properly the diverse effects caused by electroweak interactions. However, the magnitude of the anomalous muon magnetic moment calculated in terms of the theory differs from the respective experimental value. This fact indicates that the locally anisotropic fermion-antifermion condensate, rather than the scalar Higgs condensate, arises under spontaneous violation of gauge symmetry. Such a condensate gives rise to local anisotropy of space-time, thereby altering the geometric properties of the latter.

Contrary to the locally isotropic space-time described by the Riemannian geometry, the locally anisotropic space-time is described by the more general Finslerian  geometry [3]. The pioneer viable Finslerian model for space-time was constructed in [4]--[19]. In terms of the model, the symmetry with respect to local three-dimensional rotations of space proves to be partially broken, but the relativistic symmetry (i.e. symmetry with respect to transformations that relate different locally inertial reference frames to each other) remains valid, although the transformations proper differ from the Lorentz transformations.

It should be noted finally that, in addition to the relativistically-invariant Finslerian event space with partially broken 3D isotropy, works [20]--[22] described the 3-parametric family of flat relativistically-invariant Finslerian event spaces with entirely broken 3D isotropy. As to the well-known Berwald-Mo\'or space, it belongs just to that family. Considering that the properties of the Berwald-Mo\'or space are studied in detail in [23], we shall only treat the event space with partially broken 3D isotropy.

\vspace*{10mm}
\noindent
{\large{\bf 2\,. Doppler effect in the flat Finslerian event space with partially broken 3D isotropy}}
 
\vspace*{5mm}
\noindent
According to [4] and [6], the metric  of a flat Finslerian event space with partially broken 3D isotropy is
\begin{equation}\label{1}
ds^2=\left[\frac{(dx_0-\boldsymbol\nu d\boldsymbol x)^2}{dx_0^2-d\boldsymbol x^{\,2}}\right]^r
(dx_0^2-d\boldsymbol x^{\,2})\,,
\end{equation}
where the unit vector $\,\boldsymbol\nu\,$ indicates a preferred direction in 3D space while the parameter $r$ determines the magnitude of space anisotropy, characterizing the degree of
deviation of the metric (1) from the Minkowski metric. Thus, the anisotropic 
event space (1) is a generalization of the isotropic Minkowski space of 
conventional special relativity theory.

The 3-parameter noncompact group of the generalized Lorentz transformations, which link the physically equivalent inertial reference
frames in the anisotropic space-time (1) and leave
the metric (1) to be invariant, is
\begin{equation}\label{2}
{x\,'}^{\,i}=D(\boldsymbol v,\boldsymbol\nu )\,R^i_j(\boldsymbol v,\boldsymbol\nu )\,L^j_k(\boldsymbol v)\,
x^k\,,
\end{equation}
where $\boldsymbol v$ stands for the velocities of moving (primed) inertial reference frames; the matrices 
$\,L^j_k(\boldsymbol v)\,$ present the ordinary Lorentz boosts; the matrices 
$\,R^i_j(\boldsymbol v,\boldsymbol\nu )\,$ present additional rotations of the spatial axes of 
the moving frames around the vectors $\,[\boldsymbol v\,\boldsymbol\nu ]\,$ through the angles
\begin{equation}\nonumber
\varphi=\arccos\left\{ 1-\frac{(1-\sqrt{1-\boldsymbol v^{\,2}/c^2})
[\boldsymbol v\boldsymbol\nu ]^2}{
(1-\boldsymbol v\boldsymbol\nu /c)\boldsymbol v^{\,2}}\right\}
\end{equation} 
of relativistic aberration of $\boldsymbol\nu\,;$ and the diagonal matrices
\begin{equation}\nonumber
D(\boldsymbol v,\boldsymbol\nu )=\left(\frac{1-\boldsymbol v\boldsymbol\nu /c}
{\sqrt{1-\boldsymbol v^{\,2}/c^2}}
\right)^rI
\end{equation}
present the additional dilatational transformations of the event coordinates. The structure of the
transformations (2) ensures the fact that, despite of a new geometry 
of event space, the 3-velocity space remains to be a Lobachevski space.

To obtain an exact relativistic formula that would describe the Doppler effect in the flat locally anisotropic space (1), we must know how the components of the wave 4-vector $\,k^i=(\omega /c\,, \boldsymbol k)\,$ are related to each other in the initial and moving reference frames. It can be readily demonstrated (see [6] for instance) that the components of the wave 4-vector are transformed as
\begin{equation}\label{3}
{k\,'}^{\,i}=D^{-1}\,R^i_n\,L^n_j\,k^j\,;
\end{equation}
whence it follows that, under the generalized Lorentz transformations (2) and (3), the scale transformation $\,D^{-1}\,$ of the wave 4-vector is inverse to the scale transformation $\,D\,$ of event coordinates, so that the plane wave phase $\,(k^0x^0-\boldsymbol{k\,x})\,$ is an invariant of generalized Lorentz transformations.

The transformation (3) was used in [5] to obtain the relation
\begin{equation}\label{4}                                  
\omega ={\omega}\,'\,\frac{\sqrt{1-\boldsymbol v^{\,2}/c^2}}{1-\boldsymbol v\boldsymbol e/c}
\left (\frac{1-\boldsymbol v\boldsymbol\nu /c}{\sqrt{1-\boldsymbol v^{\,2}/c^2}}\right )^{\!r}\, \,.
\end{equation}
This is just the exact relativistic formula for the Doppler effect in the locally anisotropic space. With the anisotropy approaching zero $\,(\,r\to 0\,)\,$, the formula reduces to the classical formula of special relativity theory. In (4), $\,\boldsymbol v\,$ is the velocity of the moving reference frame; $\,{\omega}\,'\,$ is the frequency of a ray in this frame; $\,\omega \,$ and $\,\boldsymbol e\,$ are the frequency and direction of the ray in the initial reference frame; $\,\boldsymbol\nu\,$ is the unit vector along the preferred direction in the initial reference frame. It is seen that , as may be expected, the Doppler effect in the locally anisotropic space proves to be sensitive to the orientation of the experimental setup (
having fixed $\,{\omega}\,'\,$, $\,\vert{\boldsymbol
v}\vert\,$ and $\,\boldsymbol v\boldsymbol e\,$, we obtain the dependence of $\,\omega \,$ on the angle between $\,\boldsymbol v\,$ and $\,\boldsymbol\nu\,$ ). It is this fact that makes it possible, in principle, to detect the anisotropy of space by measuring the Doppler shift, the real possibility of such detection being limited by the magnitude of the sought anisotropy $\,(\,r\ll 1\,)\,,$ by the degree of source monochromaticity, and by the resolving power of receiver. First, consider a combination of a M\"ossbauer source and a resonant absorber.

\vspace*{10mm}
\noindent
{\large{\bf 3\,. The experiments with searching the space anisotropy by measuring the transverse Doppler effect through the M\"ossbauer effect}}
 
\vspace*{5mm}
\noindent
As long ago as during the period when the na\"ive theory of absolute ether was the only alternative to special relativity, M{\o}ller [24] suggested the experiments based on precise measurements of the Doppler effect. The respective ``null-experiments'' were expected either to detect a very weak ether wind or to become the direct and most accurate verification of special relativity.

Considering the possible local anisotropy of real space time, which is disregarded by special relativity, it would be useful to compare, up to the second order of velocities, among the values of the Doppler shift of frequencies $\Delta\,\omega
/\omega = (\omega _a-\omega _s)/\omega _s$, predicted by the pre-relativistic (PR) theory of absolute ether, by special relativity (SR), and by the relativistic theory of locally anisotropic space (AR)\,:
\begin{eqnarray}\label{5}
& &\left (\Delta\,\omega /\omega \right )^{PR} =\boldsymbol{eu}/c+(\boldsymbol{eu})(\boldsymbol e{\boldsymbol v}_s)/c^2-\boldsymbol{wu}/c^2\,, \\
& &\left (\Delta\,\omega /\omega \right )^{SR} =\boldsymbol{eu}/c+(\boldsymbol{eu})(\boldsymbol e{\boldsymbol v}_s)/c^2+({\boldsymbol v}_a^2-{\boldsymbol v}_s^2)/(2c^2)\,,\label{6}\\
& &\left (\Delta\,\omega /\omega \right )^{AR}  =\boldsymbol{eu}/c+(\boldsymbol{eu})(\boldsymbol e{\boldsymbol v}_s)/c^2+({\boldsymbol v}_a^2-{\boldsymbol v}_s^2)/(2c^2)-r\boldsymbol{\nu u}/c\,.\label{7}
\end{eqnarray}
Here, ${\boldsymbol v}_s$ and ${\boldsymbol v}_a$  are the velocities of M\"ossbauer source and absorber with respect to laboratory\,; $\boldsymbol u={\boldsymbol v}_s-{\boldsymbol v}_a$\,; $\boldsymbol e$ is unit vector along a beam in the laboratory reference frame\,; $\boldsymbol w$ is the ether wind velocity with respect to laboratory\,; $\boldsymbol\nu$ is a unit vector that indicates a locally preferred direction just where the laboratory is located in the space\,; $r$ is a local magnitude of local space anisotropy\,; $\omega _a$ is the frequency seen by the absorber\,; $\omega _s$ is the source proper frequency. Formulas (5) and (6) have been borrowed from [24], while formula (7) ensues from the exact relation (4) that describes the relativistic Doppler effect allowing for a local space anisotropy. Obviously, formulas (5)--(7) give, generally, different predictions as regards the dependence of $\,\Delta\,\omega
/\omega\,$ on the source and absorber velocities. As noted in [24], if the source and absorber are both fixed on a rotating rotor, then $\,(\boldsymbol{eu})=0\,$ with respect to the laboratory reference frame. In this case, the predictions of the theories compared are
\begin{eqnarray}\label{8}
& &\left (\Delta\,\omega /\omega \right )^{PR} =-\boldsymbol{wu}/c^2\,, \\
& &\left (\Delta\,\omega /\omega \right )^{SR} =({\boldsymbol v}_a^2-{\boldsymbol v}_s^2)/(2c^2)\,, \label{9}\\
& &\left (\Delta\,\omega /\omega \right )^{AR}  =({\boldsymbol v}_a^2-{\boldsymbol v}_s^2)/(2c^2)-r\boldsymbol{\nu u}/c\,.\label{10}
\end{eqnarray}

\vspace*{3mm} 
Let us find out now what are the results of the Doppler effect measurement experiments. In [25], the M\"ossbauer source $\,C\!o^{57}\,$ was placed at the rotor center, and the absorber $\,F\!e^{57}\,$ on the rotor radius. In that experiment, $\,v_s=0\,.$ Therefore, formulas (8)--(10) take the form
\begin{eqnarray}\label{11}
& &\left (\Delta\,\omega /\omega \right )^{PR} =\boldsymbol w{\boldsymbol v}_a/c^2\,, \\
& &\left (\Delta\,\omega /\omega \right )^{SR} ={\boldsymbol v}_a^2/(2c^2)\,, \label{12}\\
& &\left (\Delta\,\omega /\omega \right )^{AR}  ={\boldsymbol v}_a^2/(2c^2)+r\boldsymbol\nu{\boldsymbol v}_a/c\,.\label{13}
\end{eqnarray}
To increase the statistics strength, work [25] made use of two fixed counters located near the diametrically opposite sides of the rotor. Since the resonance curves were read out by summing up the readouts of both counters, the theoretical values of (11)--(13) averaged over two opposite directions of $\,{\boldsymbol v}_a\,,$ i.e. 
\begin{equation}\label{14}
\left\langle\left (\Delta\,\omega /\omega \right )^{PR}\right\rangle =0\,,
\end{equation}
\begin{equation}\label{15}
\left\langle\left (\Delta\,\omega /\omega \right )^{SR}\right\rangle = \left\langle\left (\Delta\,\omega /\omega \right )^{AR}\right\rangle ={\boldsymbol v}_a^2/(2c^2)
\end{equation}
must be compared with the Doppler shift measurement results. Such a comparison shows that the Doppler shifts measured in [25] agree, up to an error of 1.1${\%}\,,$ with the predictions (15) of special relativity and relativistic theory of anisotropic space-time, but contradict the pre-relativistic theory of absolute ether, whence, seemingly, it follows unambiguously that other experiments aimed at searching for ether wind (in particular, work [26] that, in terms of special relativity, is of null-experiment type) are senseless. Howevevr, this is not the fact, so the experiment [26] deserves particular consideration.

In [26], the Doppler shift of frequency was measured between source and absorber  placed at equal and diametrically opposite distances from the center of a rapidly rotating rotor. Since in this case $\,{\boldsymbol v}_s=-{\boldsymbol v}_a\,,$ then, according to (9) and (8), the special relativity predicts zero frequency shift, while the pre-relativistic theory predicts the effect that differs from zero\,:
\begin{equation}\label{16}
\left (\Delta\,\omega /\omega \right )^{PR} =2\boldsymbol w{\boldsymbol v}_a/c^2\,.
\end{equation}
It is remarkable that the effect similar to (16) is also predicted by the relativistic theory of anisotropic space-time. In fact, at $\,{\boldsymbol v}_s=-{\boldsymbol v}_a\,$ formula (10) gives
\begin{equation}\label{17}
\left (\Delta\,\omega /\omega \right )^{AR}  =2rc\boldsymbol\nu{\boldsymbol v}_a/c^2\,.
\end{equation}
It should be noted that it is not accidental that the light velocity $\,c\,$ is retained in the numerator of the last expression. Comparison between (17) and (16) permits $\,rc\boldsymbol\nu\,$ to be imparted the sense of vector $\,\boldsymbol w\,,$ i.e. the meaning of ether wind velocity. Thus, whereas the pre-relativistic theory of absolute ether and ether wind fails to withstand an experimental verification, the relativistic theory of anisotropic space-time revives these concepts in a sense, but on the strict relativistic basis. From the very definition of the ether wind velocity as physical quantity $\,rc\boldsymbol\nu\,$ it follows that it is the same in all physically equivalent Galilean frames. In other words, $\,rc\boldsymbol\nu\,$ is an invariant of the Lorentz transformations generalized for anisotropic space (1), i.e. is an invariant of transformations (2). It should be noted also that the same invariant enters the definition of the rest momentun ${\,\boldsymbol P}_{rest}=mrc\boldsymbol\nu\,$ that, together with the rest energy $\,E_{rest}=mc^2\,,$ characterizes a particle at rest in the anisotropic space. In view of the above, the experiment [26] has a new (active) sense and must be understood to be intended directly for searching a local space anisotropy, so the results presented in [26] must be appraised considering this fact.

\vspace*{3mm}
First of all, proceeding from [26], we shall find the restriction on the anisotropy magnitude. Let us designate the spherical (rather than geographic) coordinates on the Earth surface as $\,\Theta\,, \Phi\,,$ where $\,\Theta\,$ is the polar angle measured from the North Pole\,; $\,\Phi\,$ is the azimuthal angle measured eastwards $(\, 0\le\Theta\le\pi\, ;\, \,0\le\Phi <2\pi\, )\,.$ Let also $\,\vartheta\, $ designate the angle between a preferred direction in space $\,\boldsymbol\nu\, $ and the Earth rotation axis. Then,
\begin{eqnarray}\label{18}
& &{\nu}_{e_{_R}}(t)=\boldsymbol\nu{\boldsymbol e}_{_R}(t)=\sin\vartheta\sin\Theta\cos (\Phi +\Omega t)+\cos\vartheta\cos\Theta\,,\\
& &{\nu}_{e_{_\Theta}}(t)=\boldsymbol\nu{\boldsymbol e}_{_\Theta}(t)=\sin\vartheta\cos\Theta\cos (\Phi +\Omega t)-\cos\vartheta\sin\Theta\,,\label{19}\\
& &{\nu}_{e_{_\Phi}}(t)=\boldsymbol\nu{\boldsymbol e}_{_\Phi}(t)=-\sin\vartheta\sin (\Phi +\Omega t)\,,\label{20}
\end{eqnarray}
where $\,{\nu}_{e_{_R}}(t)\,, \,{\nu}_{e_{_\Theta}}(t)\,$ and $\,{\nu}_{e_{_\Phi}}(t)\,$ are the time- and location on earth-dependent projections of unit vector $\,\boldsymbol\nu\, $ on a surface normal, meridian, and parallel, respectively\,; $\,\Omega\,$ is angular velocity of earth rotation.

The experiment [26] was made using two fixed counters (northern and southern) located near the opposite sides of the rotor. Therefore, the measurements were taken when the absorber velocity $\,{\boldsymbol v}_a\,$ was on a geographic parallel and in virtue of (20), the equality
\begin{equation}\label{21}
\boldsymbol w{\boldsymbol v}_a=rc\boldsymbol\nu{\boldsymbol v}_a=rc{\nu}_{e_{_\Phi}}v_a=-rc\sin\vartheta\,v_a\sin (\Phi +\Omega t)=-Vv_a\sin (\Phi +\Omega t)
\end{equation}
was valid, where
\begin{equation}\label{22}
V=rc\sin\vartheta
\end{equation}
is projection of ether wind velocity on equatorial plane.\\
Considering (21), the relation (16) shows that a harmonic dependence (\,with frequency $\Omega\,$) of the Doppler shift of frequency on time of a day could be expected. At any of the experimental points, the statistics was gathered for six hours, so not more than four experimental points were read out each day, with the entire observation run lasting for a few days with an interval. Considering the doubled standard error, but a few experimental points indicated a positive effect, whereas the standard errors overlapped the expected effect at most of the points. Given this situation, the processing of the entire set of experimental data obtained during a few days has given $\,V =(\,1.6\pm{2.8}\,)$m/s\,. If $\,\boldsymbol\nu\,$ is related to the direction to the Galactic center, then, according to (22), this result, when scaled to the space anisotropy magnitude, means that $\,r = (\,1.3\pm{2.4}\,)10^{-8}\,$. Thus, the experiment [26] has failed to find any ether wind and, therefore, any local space anisotropy. As to the upper limit on the ether wind velocity obtained in 1970 [26], this limit means that, in terms of space anisotropy, $\,r < 5\times 10^{-10}\,.$ 

\vspace*{10mm}
\noindent
{\large{\bf 4\,. The planned laboratory experiment to search for space anisotropy using the present-day radio physics techniques for generation of monochromatic oscillations and detection of weak signals}}
 
\vspace*{5mm}
\noindent
Finally, let us consider another planned laboratory experiment [7] aimed at searching for space anisotropy. The experiment is based on the effect of frequency modulation of harmonic oscillations incoming to a receiver that rotates at a constant velocity about a monochromatic wave emitter. We shall proceed from the exact formula
\begin{equation}\label{23}
dt\,'=\left(\frac{1-\boldsymbol v\boldsymbol\nu /c}
{\sqrt{1-\boldsymbol v^{\,2}/c^2}}
\right)^{\!r}\sqrt{1-\boldsymbol v^{\,2}/c^2}\,\,dt\,
\end{equation}
which ensues from (1) and, in terms of relativistic theory for locally anisotropic space of events, determines the course of proper time $\,dt\,'\,$ compared with laboratory time $\,dt\,.$ Considering the smallness of the anisotropy magnitude $\,r\,,$ we obtain in the lowest order of $\,v/c\,:$
\begin{equation}\label{24}
dt\,'=\left( 1-\boldsymbol v^{\,2}/(2c^{\,2})-r\boldsymbol{v\nu}/c\right)dt\,.
\end{equation}
Since the transverse Doppler effect arises exclusively from the dependence of the receiver proper time on the receiver velocity, the straightforward way of searching for space anisotropy, i.e. for verifying formula (24), is to analyze the oscillations in the receiver that rotates at a constant velocity around the monochromatic wave source.

More specifically, the experimental design is as follows. Two receivers (1 and 2) are positioned at equal and diametrically opposite distances from an emitter of a monochromatic wave with a  frequency       
 $\,{\omega}_0\,$ and rotate with an angular frequency $\,\Omega =v/R\,$  around the emitter. Assume for simplicity that vector $\,\boldsymbol\nu\,$ lies in the rotation plane. Then, we obtain by integrating (24)\,:
\begin{equation}\label{25}
t\,'=\left(1-\frac{v^2}{2c^{\,2}}\right)t\mp \frac{rv}{c\,\Omega}\sin{\Omega t}\,,
\end{equation}
where $\,t\,$ is laboratory time\,; $\,t\,'\,$ is receiver proper time. Here and henceforth, the upper and lower signs stand for receivers 1  and 2, respectively. To within the same accuracy, we get
\begin{equation}\label{26}
t=\left(1+\frac{v^2}{2c^{\,2}}\right)t\,'\pm \frac{rv}{c\,\Omega}\sin{\left(1+\frac{v^2}{2c^{\,2}}\right)\!\Omega t\,'}\,.
\end{equation}

Let $\,\Phi (t)={\omega}_0t\,$ be the phase of oscillations incoming to a receiver as a function of laboratory time. The use of (26) gives the dependence of the phase on the receiver proper time $\,t\,'\,.$ We obtain that the oscillation frequency, $\,\omega (t\,')=d\Phi (t\,')/dt\,'\,,$  measured in the receiver proper time  is
\begin{equation}\label{27}
\omega\,(t\,')=\left(1+\frac{v^2}{2c^{\,2}}\right){\omega}_0\pm \frac{rv\,{\omega}_0}{c}\cos{\left(1+\frac{v^2}{2c^{\,2}}\right)\!\Omega t\,'}\,.
\end{equation}
Thus the receiver is affected by a frequency-modulated signal, with the modulation depth being proportional to the local space anisotropy $\,r\,.$ Detection of low-frequency oscillations ( with the frequency $\,\left(1+{v}^2/(2c^{\,2})\right)\Omega\,$ ) by the receivers in combination with subsequent subtraction of the oscillations on the rotation axis will give rise to harmonic oscillations with frequency $\,\Omega\,$ and to doubled amplitude proportional to the anisotropy magnitude $\,r\,.$

The present-day radio physics techniques for recording weak signals permit the above experiment to be realized. An autogenerator stabilized by a superconducting resonator that exhibits a narrow spectral line and low amplitude fluctuations [27],\,[28] will be used as a monochromatic source. In this case the electromotive force of a signal incoming to receivers 1 and 2 can be presented as
\begin{eqnarray}
\nonumber
u\,(t\,')\!\!&\!=\!&\!\!u_0\left\{\cos\left (1+\frac{{\Omega}^2R^{\,2}}{2\,c^{\,2}}\right )\!{\omega}_0t\,'\pm \frac{r{\omega}_0R}{2\,c}\cos\left (1+\frac{{\Omega}^2R^{\,2}}{2\,c^{\,2}}\right )\!({\omega}_0+\Omega )\,t\,'\right.\\ 
\!\!&\!\!&\!\!\mp \,\left. \frac{r{\omega}_0R}{2\,c}\cos\left (1+\frac{{\Omega}^2R^2}{2c^2}\right )\!({\omega}_0-\Omega )\,t\,'\,\right\}\,.\label{28}
\end{eqnarray}
The experimental design expects a signal, which is the electromotive force component $\,\Delta {\tilde u}_s\,$ in the receiver input, whose frequency is tuned off to the left or right from the base frequency\,:
\begin{equation}\label{29}
\Delta {\tilde u}_s=\pm u_0\,\frac{r{\omega}_0R}{2\,c}\cos\left (1+\frac{{\Omega}^2R^{\,2}}{2\,c^{\,2}}\right )\!({\omega}_0+\Omega )\,t\,'\,.
\end{equation}
Equating $\,\Delta {\tilde u}_s/u_0\,$ to the relative value $\,(\Delta u/u)_{\alpha}\,$ of the natural amplitude fluctuations of autogenerator at frequency $\,{\omega}_0+\Omega\,,$ we obtain a boundary for the least detectable anisotropy $\,r_{min}\,$ determined by the source fluctuations\,:
\begin{equation}\label{30}
r_{min}=\frac{c}{R\Omega}\sqrt{\frac{kT^A_N}{WQ^2\tau}}\,\,, 
\end{equation}
where $\,T^A_N\,$ is the noise temperature of the active element of autogenerator\,, $\,Q\,$ is the quality factor of a stabilizing superconducting resonator\,, $\,W\,$ is autogenerator power and $\,\tau\,$ is measurement time. Substituting $\,R=25\,$cm, $\,\Omega =10^2\,$rad/s\,\, and the real autogenerator parameters $\,T^A_N=600\,$K, $\,W=10^{-2}\,$wt, $\,Q=2\times 10^{8}\,$ in (30), we obtain $\,r_{min}=5\times 10^{-11}/\sqrt{\tau}\,.$

The required receiver sensitivity when recording $\,\Delta {\tilde u}_s\,$ is determined from the condition
\begin{equation}\label{31}
{\left(\frac{\Delta {\tilde u}_s}{u_0}\right)}^{\!2}=\frac{\Delta W_s}{W}=\frac{r^{\,2}_{min}\,{\omega}^{\,2}_{\,0}R^{\,2}}{4\,c^{\,2}}\,\,,
\end{equation}
where $\,\Delta W_s=kT^{Rec}_N/\tau\,$ is the least detectable signal power at the receiver input\,, $\,{\omega}_0\,$ is the frequency of auto-oscillations  of source\,, $\,T^{Rec}_N\,$ is the receiver noise temperature.

At $\,{\omega}_0=2\times 10^{10}\,$rad/s \,\,and\, $\,W=10^{-2}\,$wt\,\,, expression (31) gives $\,T^{Rec}_N=10^2\,$K\,, which is an order higher than the $\,{T^{Rec}_N}\,\,$ value reached at present in the given frequency range.

\vspace*{10mm}
\noindent
{\large{\bf 5\,. Conclusion}}
 
\vspace*{5mm}
\noindent
On examining the experiments aimed at searching for ether wind that were carried out soon after the discovery of the M\"ossbauer effect, we have concluded that the experiments have given the upper limit $\,r < 5\times 10^{-10}\,$ on the space anisotropy magnitude. The present day use of the radically new rotors ($\,n\ge 6\times 10^{5}\,$turns/min\,) developed by the ITEP team (Moscow), as well as of the M\"ossbauer sources with a much narrower line width, has made it possible to lower the minimum detectable value of space anisotropy by at least three orders. Accordingly, the experiment [26] is very topical to repeat. At the same time, it should be kept in mind that, in case a new experiment of that type is aimed at searching for space anisotropy originating from the Galaxy, the northern and southern counters are expedient to replace by western and eastern counters because, according to (19), the daily average value of projection of $\,\boldsymbol\nu\,$ onto the laboratory floor equals $\,\cos\vartheta\sin\Theta\,$ and is directed northwards.

As to the experiment described above in section 4, which is expected to base on the effect of harmonic oscillation frequency modulation, the expression (30) shows that the potentiality of such an experiment in detecting a local space anisotropy is restricted solely by the operation period of a respective facility. In any case, the realization of the experiment will make it quite possible to either discover the expected space anisotropy or lower its upper boundary down to $\,\sim 10^{-14}\,.$

Finally, it should be noted that, beside the relativistic Finslerian approach, an alternative approach to the space anisotropy problem is being developed actively. The respective string-motivated theory is known to be the Extended Standard Model of strong, weak, and electromagnetic interactions, or the Standard Model Extension (SME) (see [29]--[33] for instance ). Since SME is not a relativistically-invariant theory, the anisotropy impact on the fundamental field dynamics is described by SME in terms of much more numerous parameters as compared with the set of parameters that characterize the space anisotropy proper. As a result, the complete anisotropy measurements involve measuring the very numerous effects predicted by SME. At present, attempts are made to detect and measure the respective effects by such collaborations as, for instance, \,LSND\,, KTeV\,, FOCUS\,, BaBar\,, BELLE\,, OPAL\,, DELPHI\,, and BNL\,g-2\,.
 
Contrary to the effects predicted by SME, the Doppler shift is a pure kinematic effect, while the space anisotropy impact on that effect is determined only by the parameters of the anisotropy proper. Therefore, the above described experiments are most expedient as regards searching and measuring the local space anisotropy. Besides, the realization of the experiments does not involve much material cost.

\end{document}